\documentclass[twocolumn]{aastex63}

\usepackage[utf8]{inputenc} 
\usepackage{amsmath, amsbsy}
\usepackage{hyperref} 

\shorttitle{Bar--disk chemodynamics with present and future data}
\shortauthors{Wheeler et al.}

\newcommand{\gaia}{\emph{Gaia}}
\newcommand{\gaiarv}{\emph{Gaia} radial velocity}

\newcommand{\lamost}{LAMOST}

\newcommand{\apogee}{APOGEE}

\newcommand{\jphijr}{$J_\phi$--$J_R$}
\newcommand{\rvphi}{$R$--$v_\phi$}
\newcommand{\slowbar}{\emph{slow bar}}
\newcommand{\intermediatebar}{\emph{intermediate bar}}
\newcommand{\fastbar}{\emph{fast bar}}

\newcommand{\teff}{T_\mathrm{eff}}
\newcommand{\logg}{\log(g)}

\newcommand{\T}{\intercal}
\newcommand{\kms}{\ensuremath{\mathrm{km~s^{-1}}}}
\newcommand{\kpc}{\ensuremath{\mathrm{kpc}}}
\newcommand{\kpckms}{\ensuremath{\mathrm{kpc~km~s^{-1}}}}
\newcommand{\kmskpc}{\ensuremath{\mathrm{km~s^{-1}~kpc^{-1}}}}
\newcommand{\Op}{\ensuremath{\Omega_\text{p}}}

\newcommand{\reviewer}[1]{#1}
\newcommand{\secondrevision}[1]{#1}

\begin{document}
\title{Chemodynamical signatures of bar resonances in the Galactic disk: current data and future prospects}

\author[0000-0001-7339-5136]{Adam Wheeler}
\affiliation{Department of Astronomy, Columbia University, Pupin Physics Laboratories, New York, NY 10027, USA}
\author[0000-0003-3230-4589]{Irene Abril-Cabezas}
\author[0000-0002-2806-3738]{Wilma H. Trick}
\affiliation{Max-Planck-Insitut f\"{u}r Astrophysik, Karl-Schwarzschild-Str. 1, D-85748 Garching b. M\"{u}nchen, Germany}
\author[0000-0002-0897-3013]{Francesca Fragkoudi}
\affiliation{European Southern Observatory, Karl-Schwarzschild-Str. 2, D-85748 Garching b. M\"{u}nchen, Germany}
\affiliation{Max-Planck-Insitut f\"{u}r Astrophysik, Karl-Schwarzschild-Str. 1, D-85748 Garching b. M\"{u}nchen, Germany}
\author[0000-0001-5082-6693]{Melissa Ness}
\affiliation{Department of Astronomy, Columbia University, Pupin Physics Laboratories, New York, NY 10027, USA}
\affiliation{Center for Computational Astrophysics, Flatiron Institute, 162 Fifth Avenue, New York, NY 10010, USA}
\correspondingauthor{Adam Wheeler}
\email{a.wheeler@columbia.edu}

\begin{abstract}
\noindent
The Galactic disk exhibits complex chemical and dynamical substructure thought to be induced by the bar, spiral arms, and satellites.
Here, we explore the chemical signatures of bar resonances in action and velocity space and characterize the differences between the signatures of corotation and higher-order resonances using test particle simulations.
Thanks to recent surveys, we now have large datasets containing metallicities and kinematics of stars outside the solar neighborhood.
We compare the simulations to the observational data from \gaia{} EDR3 and \lamost{} DR5 and find weak evidence for a slow bar with the ``hat'' moving group \reviewer{($250~\kms \lesssim v_\phi \lesssim 270~\kms$)} associated with its outer Lindblad resonance and ``Hercules'' \reviewer{($170~\kms \lesssim v_\phi \lesssim 195~\kms$)} with corotation.
While constraints from current data are limited by their spatial footprint, stars closer in azimuth than the Sun to the bar's minor axis show much stronger signatures of the bar's outer Lindblad and corotation resonances in test particle simulations.
Future datasets with greater azimuthal coverage, including the final \gaia{} data release, will allow reliable chemodynamical identification of bar resonances. 
\end{abstract}

\section{Introduction} \label{sec:intro}
\gaia{} \citep{gaiacollaborationGaiaMission2016}, especially data release 2 (DR2; \citealp{gaiacollaborationGaiaDataRelease2018a}), which included radial velocities, has revealed the Galactic disk to be dynamically complex \citep{gaiacollaborationGaiaDataRelease2018b, antojaDynamicallyYoungPerturbed2018}. 
Many perturbers are at play, and the specific influence of each is not yet clear.
The Galaxy's bar (e.g. \citealp{huntOuterLindbladResonance2018a, fragkoudiChemodynamicsBarredGalaxies2019}), spiral structure \citep{quillenSpiralArmCrossings2018, kawataRadialDistributionStellar2018,huntTransientSpiralStructure2018, antojaDynamicallyYoungPerturbed2018,sellwoodDiscriminatingTheoriesSpiral2019, khannaGALAHSurveyGaia2019, khoperskovHicSuntDracones2020}, and satellite galaxies \citep{laporteFootprintsSagittariusDwarf2019, khannaGALAHSurveyGaia2019} all affect disk kinematics.
Furthermore, \citet{huntSignaturesResonancePhase2019} showed that the signatures of each are difficult to disentangle and that many interpretations of the data are possible. \citet{hilmiFluctuationsGalacticBar2020} showed that perturbers can interact significantly, further complicating matters.

In this work, we focus on the chemical signatures of the bar in action space, and demonstrate that chemical abundances can aide the identification of bar resonances in the disk and constrain the bar pattern speed, $\Omega_\text{p}$.
Because most abundances do not change over a star's lifetime, they trace its birth location, providing a non-dynamical ``memory'', that can help distinguish the effects of disk perturbations.
\reviewer{ However, stars on orbits in resonance with the bar are not uniformly distributed in azimuth.
The spatial selection function and limited footprint of current data sets therefore strongly shapes the observed and expected abundance signatures, as we will discuss in this work.}
Past work on chemistry and dynamics has examined the distribution of metallicities and ages near the Sun or throughout the disk to constrain radial migration (e.g. \citealt{roskarRidingSpiralWaves2008a, schonrichChemicalEvolutionRadial2009, loebmanGenesisMilkyWay2011, haydenCHEMICALCARTOGRAPHYAPOGEE2015, martinez-medinaRevealingSpiralArms2016, loebmanImprintsRadialMigration2016a, khoperskovEscapeesBarResonances2019, frankelKeepingItCool2020}).
Recently, \citet{chibaTreeringStructureGalactic2021} examined the chemical signature of a decelerating bar.

The existence of the Milky Way's bar was first inferred from gas velocities \citep[e.g.][]{devaucouleursInterpretationVelocityDistribution1964, cohenObservationsOHGalactic1976, lisztGasDistributionCentral1980}, then with photometry \citep{blitzDirectEvidenceBar1991, weilandCOBEDiffuseInfrared1994} and star counts \citep[e.g.][]{nakadaBulgeOurGalaxy1991, whitelockShapeBulgeIRAS1992, weinbergDetectionLargescaleStellar1992}.
Over time, its pitch angle has been constrained, with the modern consensus being that the bar is ahead of the Sun by roughly $27^\circ$ \citep{dwekMorphologyNearinfraredLuminosity1995, binneyPhotometricStructureInner1997, stanekModelingGalacticBar1997, lopez-corredoiraBoxyBulgeMilky2005,  rattenburyProperMotionDispersions2007, caoNewPhotometricModel2013, weggMappingThreedimensionalDensity2013}. 
The bar's pattern speed, however, is still disputed.
Identifying the Hercules moving group, a feature of the local velocity distribution, with the Outer Linblad Resonance yields a pattern speed of roughly $55~\kmskpc$ (\citealp{dehnenPatternSpeedGalactic1999, dehnenEffectOuterLindblad2000}, later \citealp{chakrabartyPhaseSpaceStructure2007, minchevNewConstraintsGalactic2007, minchevLowvelocityStreamsSolar2010, antojaConstraintsGalacticBar2014, fragkoudiRidgesUndulationsStreams2019a}).
On the other hand, studies of the inner Galaxy suggest that the bar is slower, with a pattern speed of 30--40 $\kmskpc$ \citep{weinerPropertiesGalacticBar1999, rodriguez-fernandezGasFlowModels2008, longMadetomeasureGalaxyModels2013, weggStructureMilkyWay2015, sormaniGasFlowBarred2015}, and that Hercules could be due to the corotation (CR) resonance \citep{perez-villegasRevisitingTaleHercules2017,monariTracingHerculesGalactic2019, monariSignaturesResonancesLarge2019,binneyTrappedOrbitsSolarneighbourhood2020}.
Recent studies suggest a figure around the upper end of this range, \reviewer{and all but rule out a faster pattern speed} \citep{portailDynamicalModellingGalactic2017, clarkeMilkyWayBar2019, sandersPatternSpeedMilky2019, monariSignaturesResonancesLarge2019, bovyLifeFastLane2019, morenoEffectOrbitalTrapping2021, kawataGalacticBarResonances2021}.


While the true potential of the Milky Way is complex, asymmetric, and time-dependent, we can use an axisymmetric approximate potential model to calculate ``axisymmetric actions extimates'', $J_\phi, J_R, J_z$, for each star, associated with azimuthal, radial, and vertical motion, respectively \citep[e.g.][chapter 3]{binneyGalacticDynamicsSecond2008}.
From here on, we refer to axisymmetric action estimates simply as \emph{actions}, \reviewer{which are always calculated using axisymmetric potentials, even when applied to non-axisymmetric simulations.}
Unlike true actions, which are conserved but not generally known, these change as the star orbits in a non-axisymmetric potential \citep[e.g.][]{lynden-bellGeneratingMechanismSpiral1972, sellwoodRadialMixingGalactic2002, binneyOrbitalToriNonaxisymmetric2018, trickIdentifyingResonancesGalactic2021}.
\secondrevision{
The fact that they are not true actions does not pose a problem in this context, nor does this generate a mismatch between simulations and data
Our axisymmetric action-angles provide a valid coordinate system for the orbits, that is analogous to other reference frames, for example $(R,\phi,z,v_R,v_\phi,v_z)$ or $(\mathrm{ra},\mathrm{dec},d,\mathrm{pm}_\mathrm{ra},\mathrm{pm}_\mathrm{dec},v_\mathrm{los})$. 
}
Stars on orbits resonant with the bar and with small $J_z$ librate around negatively-sloped lines in the $J_R$--$J_\phi$ plane, with $J_R$ decreasing with increasing $J_\phi$ along them \citep{sellwoodRecentLindbladResonance2010}. 
We refer to these as axisymmetric resonance lines.

The primary way the bar impacts the disk is through resonant trapping, which affects stellar orbits with frequencies commensurate with the bar, i.e. for which
\begin{equation} \label{eq:res}
    m(\Op - \Omega_\phi) = l \Omega_R \quad ,
\end{equation}
where $m$ and $l$ are small integers with $m > 0$, and $\Omega_\phi$ and $\Omega_R$ are the frequencies of azimuthal and radial oscillation (approximately equal to the frequencies following from the axisymmetric Hamiltonian), respectively.
These orbits librate around their parent orbits, which are closed in the frame that rotates with the bar \citep{contopoulosOrbitsBarredGalaxies1989, athanassoulaMorphologyBarOrbits1992}.  
Because of the conservation of the Jacobi integral, changes to the axisymmetric actions of stars obey 
\begin{equation}\label{eq:jjlm}
    \Delta J_R = \frac{l}{m} \Delta J_\phi
\end{equation}
to first order \citep{lynden-bellGeneratingMechanismSpiral1972, sellwoodRadialMixingGalactic2002}.
Consequentially, corotating (0:1, i.e. $l=0$, $m=1$) stars can undergo changes to their angular momentum, $J_\phi$, without being radially heated.
Most stars are on near-circular orbits, meaning that scattering at the outer 1:1, 1:4, and 1:2 Linblad resonances preferentially increase eccentricity, $J_R$, inducing a net outward migration and leading to under- and over-densities on either side of the axisymmetric resonance line.
See \citet{trickIdentifyingResonancesGalactic2021} for a pedagogical introduction and numerical investigation of these effects.


In sections \ref{sec:data} and \ref{sec:sims} we describe the observational data and simulations we use, respectively.
In section \ref{sec:results}, we show how bar resonances give rise to distinct chemical signatures and compare them to observations.
In section \ref{sec:discussion}, we discuss the impact of spatial selection functions and consider the capabilities of future surveys.
We conclude in section \ref{sec:conclusions}.

\section{Data}\label{sec:data}
We use astrometry from \gaia\ EDR3 \citep{gaiacollaborationGaiaEarlyData2020}, which includes positions, parallaxes, and proper motions for $1.3 \times 10^9$ sources, as well as radial velocities \reviewer{for $7.2 \times 10^7$ stars} (hereafter the \gaiarv\ sample; these are the same as in DR2).
We also use the \lamost\ \citep{cuiLargeSkyArea2012, zhaoLAMOSTSpectralSurvey2012, dengLAMOSTExperimentGalactic2012} DR 5 FGAK sample, which provides radial velocities and metallicities for $4.2 \times 10^6$ stars.
We use distance estimates from \citet{bailer-jonesEstimatingDistancesParallaxes2018}, which used a Galactic prior together with \gaia\ parallax to obtain distance posteriors for stars from both catalogs.

\begin{figure}
    \centering
    \includegraphics[width=0.45\textwidth]{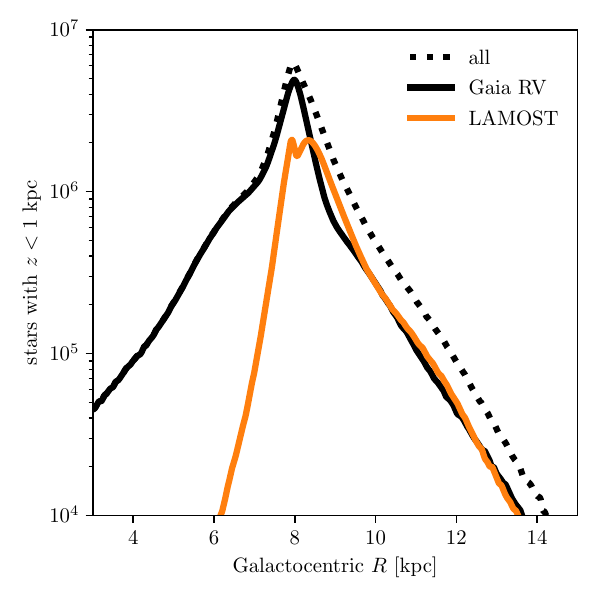}
    \caption{The radial distribution of stars within 1 $\kpc$ from the Galactic midplane in each input catalog, as well as in the crossmatched joint catalog.
    \lamost\ radial velocities can fruitfully augment \gaia\ in the outer disk.}
    \label{fig:footprints}
\end{figure}

Figure \ref{fig:footprints} shows the radial distribution of \lamost{} and \gaiarv{} stars.
\reviewer{Interior to the Sun, there  are  far  more \gaiarv{} than \lamost{} stars, and future \gaia{} data releases will include more radial velocity measurements.
But at present, \lamost{} has observed a similar number of stars at $R$ beyond the Sun, many of which where not observed by \gaiarv{}. 
Beyond $R=10~\kpc$, \lamost{} nearly doubles the number of stars with radial velocities available.
Despite the fact that the \lamost{} has a larger typical radial velocity uncertainty ($\sim5~\kms$, as opposed to \gaia's $\sim1~\kms$), it usefully augments \gaia{} in the outer disk.
}

We used \texttt{galpy} \citep{bovyGalpyPythonLibrary2015} to calculate actions for each star with the St\"{a}ckel fudge \citep{binneyActionsAxisymmetricPotentials2012, bovyDirectDynamicalMeasurement2013} in the axisymmetric \texttt{MWPotential2014} model \citep{bovyGalpyPythonLibrary2015}.
\reviewer{The \texttt{MWPotential2014} assumes the solar Galactocentric radius to be $R_0 = 8~\kpc$ and the local standard of rest to be $v_\text{circ}(R_0) = 220~\kms$. While more recent MW models suggest slightly different values (e.g. \citealp{gravitycollaborationDetectionGravitationalRedshift2018}), they do not have a strong impact on our results. This is illustrated in Appendix \ref{appendix:eilers}, where we show our central observational plot with actions calculated using the potential from \citet{eilersCircularVelocityCurve2019}.
For the solar proper motion, we use $v_X = -11.1~\kms$, $v_Y = 232.24~\kms$, and $v_z = 7.25~\kms$ by \citet{schonrichChemicalEvolutionRadial2009}.}
While kinematic uncertainties propagate to actions, they are small enough not to obscure the chemodynamical signatures we investigate in this paper.
\citet[Appendix D]{wheelerAbundancesMilkyWay2020} show, for example, mean action uncertainties as a function of Galactocentric radius, $R$, for \gaia{} DR2 and \lamost{} DR 4, which have radial velocity uncertainties very similar to \gaia{} EDR3 and \lamost{} DR 5.
\reviewer{In addition, we demonstrate in Appendix \ref{appendix:fuzz} that our results stay robust when applying simulated observational errors to one of our test-particle simulations.}

In order to focus the investigation on near-circular in-plane disk orbits, we used only stars with $J_z < 10~\kpckms$ and $|z| < 1~\kpc$.
This selection is dominated by thin disk, low-alpha sequence stars.
We also removed the roughly $1\%$ of stars with $L_z$ outside $1000 - 3000~\kpckms$ to facilitate comparison with simulations.
In order to mitigate magnitude-driven selection effects, we use only stars with $2.2 < \log g < 2.7$ and $4500~\mathrm{K} < T_\mathrm{eff} < 5000~\mathrm{K}$ when dealing with abundances, which often have stellar-parameter-dependent systematics.
This selection corresponds roughly to the red clump.
\reviewer{These cuts leave roughly $2 \times 10^6$ stars when abundances are not needed, and $9 \times 10^4$ when they are.
For stars in \lamost{} and \gaiarv{}, the radial velocity with the lower uncertainty is used (this is \gaiarv{} for about 99\% of stars).  
The mean radial velocity uncertainty for the sample with all cuts, including those in $\teff$ and $\logg$, is $3.2~\kms$  , with roughly half of radial velocities coming from \gaia{} and from \lamost{}}.

\subsection{The MW disk as seen in this dataset} \label{sec:dataoverview}
\begin{figure*}
    \centering
    \includegraphics[width=\textwidth]{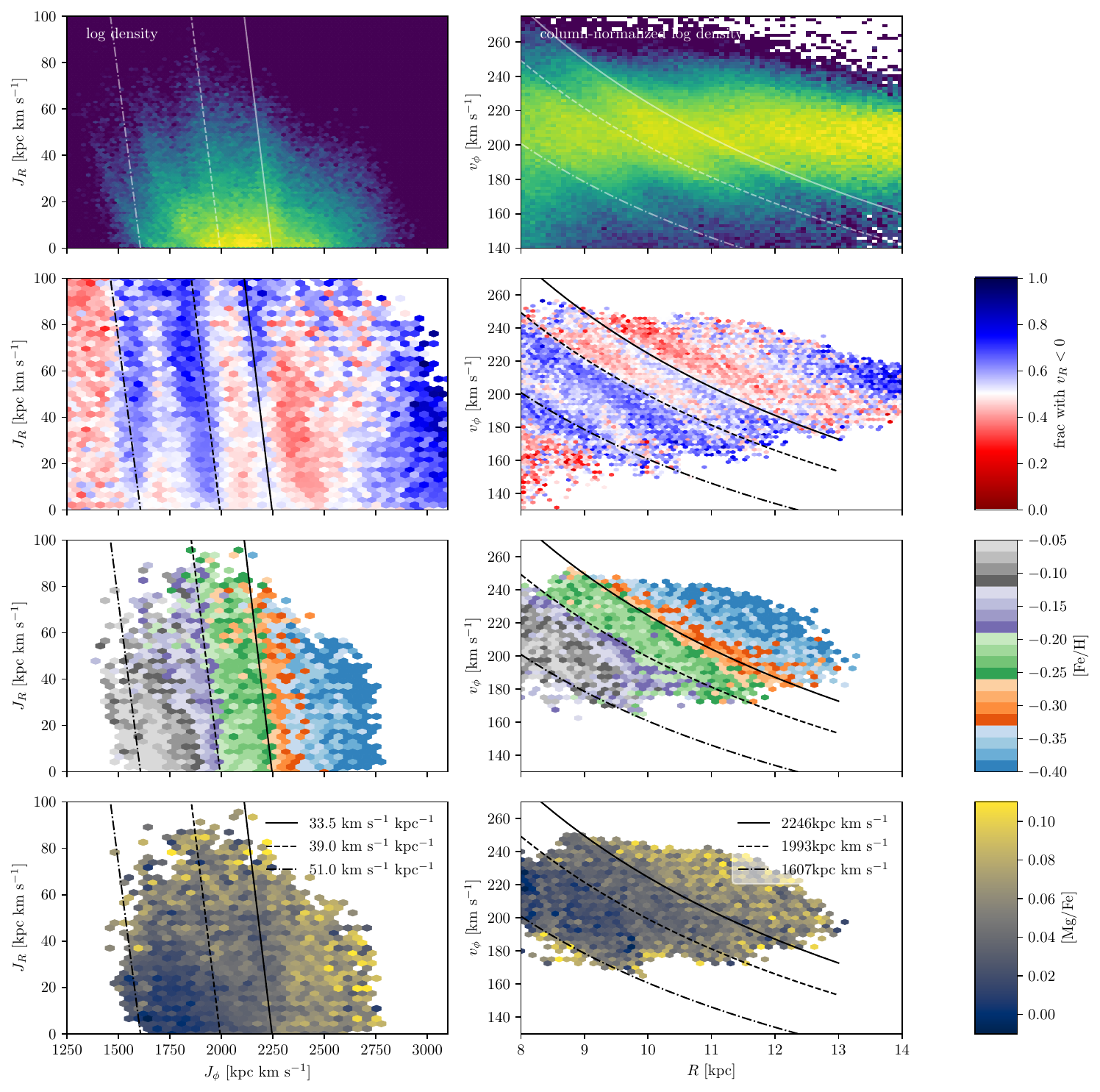}
    \caption{The \lamost{} \jphijr{} and \rvphi{} planes in stellar density (top row; $\sim10^6$ stars), fraction of inwards-moving stars (second row; $\sim10^6$ stars), mean [Fe/H] (third row; $\sim10^5$ stars) and mean [Mg/Fe] (bottom row; $\sim10^5$ stars).
    The OLR axisymmetric resonance lines for the three pattern speeds we consider are overplotted on the \jphijr{} plane, and contours of constant $J_\phi$ corresponding the circular OLR orbits for the same pattern speeds are overplotted on the \rvphi{} plane.
    The top four panels contain $~800,000$ stars and the bottom four contain $~95,000$, because we made additional cuts to mitigate abundance systematics.
    We use fewer stars when plotting metallicities or abundances because we restrict our sample to a small part of $\logg$--$\teff$ space to minimize systematics.}
    \label{fig:lamost_kinematics}
\end{figure*}
   
Figure \ref{fig:lamost_kinematics} shows the \gaia{} + \lamost{} \rvphi{} and \jphijr{} planes in stellar density, fraction of inwards-moving stars, mean [Fe/H] and mean [Mg/Fe], which is taken from \citet{wheelerAbundancesMilkyWay2020}, who estimated detailed abundances for \lamost{} DR4.
Note that [Fe/H] and [Mg/Fe] increase and decrease, respectively, with increasing $J_\phi$, $J_R$, $R$, and $v_\phi$
While we make use of [Fe/H] only in the remainder of this paper, we note that the results presented here can in principle be generalized to other abundances.
The axisymmetric resonance lines for the outer Lindblad resonance (OLR) of $\Op = 33.5~\kmskpc$, $39~\kmskpc$, and $51~\kmskpc$ are plotted for orientation. In the \mbox{\rvphi{}}
plane, contours of constant angular momentum are overplotted, corresponding to the values of $J_\phi$ on these OLR axisymmetric resonance lines at $J_R = 0$.
These pattern speeds, used throughout the paper, correspond to a \fastbar{} whose OLR is associated with Hercules ($51~\kmskpc$), an \intermediatebar{} ($39~\kmskpc)$, and a \slowbar{} ($33.5~\kmskpc$).
The \intermediatebar{} and \slowbar{} OLR axisymmetric resonance lines align with the ``Sirius'' and ``hat'' ridges, respectively, in the $J_\phi$ -- $J_R$ plane.
The \fastbar{} OLR axisymmetric resonance line is slightly to the left of the inward-moving ridge that contains the "Horn" stars, and to the right of the outward-moving ``Hercules'' ridge with which the OLR of a rapidly rotating bar is often associated.
The exact values of the \slowbar{} and \fastbar{} pattern speeds are chosen so that their OLR axisymmetric resonance lines align with transitions between net inwards and outwards movement with increasing $J_\phi$, as proposed by \citet{trickIdentifyingResonancesGalactic2021}. 

In the \rvphi{} plane, the ridges identified  with \gaia{} \citep{kawataRadialDistributionStellar2018, gaiacollaborationGaiaEarlyData2021} and attributed variously to the bar, spiral arms, the Sagittarius dwarf galaxy, or a combination thereof \citep{antojaDynamicallyYoungPerturbed2018, khannaGALAHSurveyGaia2019, bland-hawthornGALAHSurveyGaia2019, fragkoudiChemodynamicsBarredGalaxies2019, fragkoudiRidgesUndulationsStreams2019a, laporteFootprintsSagittariusDwarf2019} are clearly visible.
The largest of these can be seen extending to a larger Galactocentric radius, $R$, than with \gaia{} data alone.

\section{Simulations} \label{sec:sims}
To better understand the connection between bar resonance and chemistry in action space, we use test particle simulations with the same pattern speeds introduced in section \ref{sec:dataoverview} ($\Op = 33.5~\kmskpc$, $39~\kmskpc$, and $51~\kmskpc$). The setup of the simulations is analogous to those used by  \citet{trickIdentifyingResonancesGalactic2021} and \citet{trickIdentifyingResonancesGalactic2022}, starting with an axisymmetric disk \citep{binneyModelsOurGalaxy2011} of massless particles, followed by integrating stellar orbits in an analytic MW potential model \citep{bovyGalpyPythonLibrary2015} that slowly introduces a bar perturbation \citep{dehnenEffectOuterLindblad2000, monariEffectsBarspiralCoupling2016}. The boxy bar model has $m=2$ and $m=4$ components \citep{huntOuterLindbladResonance2018a} with strengths $\alpha_{m=2} = 0.01$ and weak $\alpha_{m=4} = -0.0005$. 
Orbits are integrated for 50 bar periods and the simulation snapshots at 20, 30, 40, and 50 bar periods are stacked to increase the number of particles to $4\times10^7$.
\reviewer{Axisymmetric approximate actions are calculated exactly as for the observational data (see Section \ref{sec:data} for details).}

We assign metallicity values to test particles based on their positions in the intial disk, with a metallicity decreasing with both Galactocentric cylindrical radius $R$, and height from the midplane, $|z|$.
Each particle was assigned an iron abundance according to
\begin{equation} \label{eq:feh}
    \mathrm{[Fe/H]} = -0.09~\left(\frac{R}{\kpc} - 6\right) + 0.2 - 0.21 \frac{|z|}{\kpc} + \varepsilon \quad ,
\end{equation}
where $\varepsilon$ is a random Gaussian perturbation with standard distribution 0.1, \reviewer{roughly the mean uncertainty of the LAMOST [Fe/H] values.}
The $R$ and $z$ dependence of this relation are based on the trends observed by \citet{haydenChemicalCartographyAPOGEE2014} in the outer low-alpha disk.
\reviewer{We note that the qualitative results of this work do not depend on the details of how metallicities are assigned, so long as the dominant trend is decreasing metallicity with Galactocentric radius (in Appendix \ref{appendix:alternate_metallicity} we show that assigning metallicities based on initial action does not change our results).}

\section{Results} \label{sec:results}

\subsection{Metallicity signatures of bar resonances in action space} \label{sec:explanation}
\begin{figure*}
    \centering
    \includegraphics[width=0.95\textwidth]{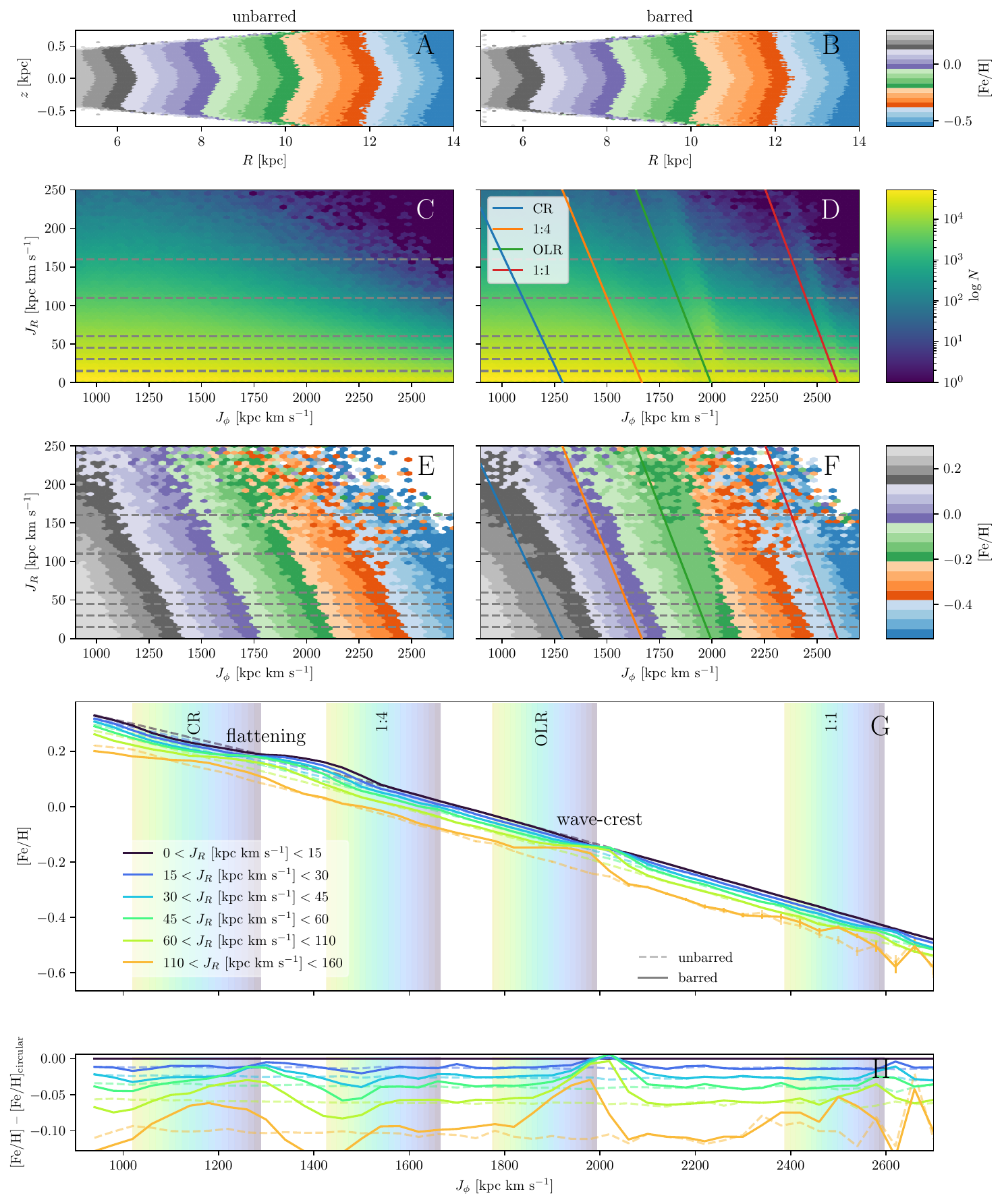}
    \caption{The metallicity signatures at the bar resonances in the \intermediatebar{} test particle simulation.
    \textbf{A and B:} mean metallicity in the meridional plane for the axisymmetric and barred test particle simulations, respectively.
    \textbf{C and D:} number density in the \jphijr{} plane for the axisymmetric and barred test particle simulations, respectively.
    \textbf{E and F:} mean [Fe/H] in the \jphijr{} plane for the axisymmetric and barred test particle simulations.
    \textbf{G:} mean [Fe/H], in bins of $J_\phi$ and $J_R$, for the axisymmetric (dashed) and \intermediatebar{} (solid) test particle simulations.
    The CR resonance flattens the mean-metallicity trends, while the higher-order resonances (especially the OLR) create wave-crest signatures.
    \textbf{H:} differences between the above: the difference between mean [Fe/H] of each $J_R$ bin and the lowest $J_R$ bin.
    Error bars in the bottom two panels show the standard error in each bin.
    Except at the largest values of $J_R$ and $J_\phi$, the standard errors are too small to be visible, indicating that the trends are robust.
    The colored bands show the range of $J_\phi$ values each axisymmetric resonance line takes on below $160~\kpckms$.
    See also Appendix \ref{appendix:fuzz}.
    }
    \label{fig:irene_explanation}
\end{figure*}

While bar resonances can change $J_\phi$ and $J_R$, they can not change a star's atmospheric metallicity, which remains very-nearly constant over a star's lifetime.
As discussed in Section \ref{sec:intro}, actions oscillate in specific proportions at resonances, so kinematically distinguishing between resonances in observational data is possible in principle. 
Stellar metallicities, however, provide an orthogonal source of information that makes this task much easier. 
Figure \ref{fig:irene_explanation} demonstrates how bar resonances lead to signatures in $J_\phi$--[Fe/H] trends at different $J_R$.
The left column shows test particles evolved in an axisymmetric potential and the right shows particles evolved in the barred potential discussed in Section \ref{sec:sims}.
Panels A and B show the $R$--$z$ plane in metallicity, and panels C-F show the \jphijr{} plane in both metallicity and density.
The bottom two panels show the [Fe/H] trends as a function of $J_\phi$, in bins of $J_R$ (panel G shows absolute metallicities, and panel H shows differences with respect to the smallest $J_R$ bin).
The error bars in the bottom two panels show the standard error, the scatter in metallicity over the square root of number of stars in each bin.
\reviewer{Except at the largest values of $J_R$ and $J_\phi$, the standard errors are too small to be visible, indicating that the trends are not due to sampling errors and that the effect of $\varepsilon$ in (\ref{eq:feh}) has been washed out.
Unbiased measurement error in the observational data will be similarly washed out.}
The colored bands in panels G and H show the range of $J_\phi$ values each axisymmetric resonance line takes on below $160~\kpckms$.
The $J_R$ bin boundaries used in panels G and H are shown as horizontal lines in panels C-F.

Because stars in corotation with the bar oscillate strongly in $J_\phi$ without changing coherently in $J_R$  (Equation \ref{eq:jjlm}), and density changes slowly with $J_\phi$, the effect of the corotation resonance is difficult to see in stellar density in the \jphijr{} plane.
Contrasting the mean [Fe/H] around the CR axisymmetric resonance line for the barred and unbarred simulations reveals that the metallicity gradient becomes shallower when the bar causes stars to librate symmetrically in $J_\phi$, washing out the initial metallicity gradient.
This flattening is easily visible in panel G, where the metallicity gradient is visibly flattened.

At the other resonances (OLR, 1:1, and 1:4), $J_R$ changes in lockstep with $J_\phi$ (equation \ref{eq:jjlm}).
Because of the strong density gradient in $J_R$ (and weaker gradient in $J_\phi$), more stars are on average displaced from near-circular, low-$J_R$ orbits to eccentric, high-$J_R$ orbits than vice versa, creating a density ridge to the right of the corresponding axisymmetric resonance line (panel D). 
Below approximately $J_R = 50~\kpckms$, the metallicity contours to the right of the OLR axisymmetric resonance line are more vertical in the barred simulation because higher-metallicity stars on circular orbits have moved upward (and slightly to the right) in the \jphijr{} plane, shifting the mean metallicity up at the location of the overdensity ridge.
In panels G and H, the chemical trends around the OLR are most easily identifiable.
For near-circular orbits ($0 < J_R~[\text{kpc km s}^{-1}] < 15$) in the barred potential, the mean metallicity drops slightly below that in the unbarred potential.
This is because of the small number of lower-metallicity stars scattered from larger $J_\phi$ and $J_R$ to more circular orbits.
In contrast, among the high-$J_R$ stars, the deviation from the overall metallicity trend rises above that of the unbarred potential to form a ``wave-crest'' in the $J_\phi$-[Fe/H] plane across bins of $J_R$. 
Like the OLR, the 1:1 and 1:4 resonances create metallicity wave-crests, but they are weaker because of the weaker Fourier $m = 4$ bar component, and the 1:1 comprising stars far beyond the bar.

\begin{figure}
    \centering
    \includegraphics[width=0.48\textwidth]{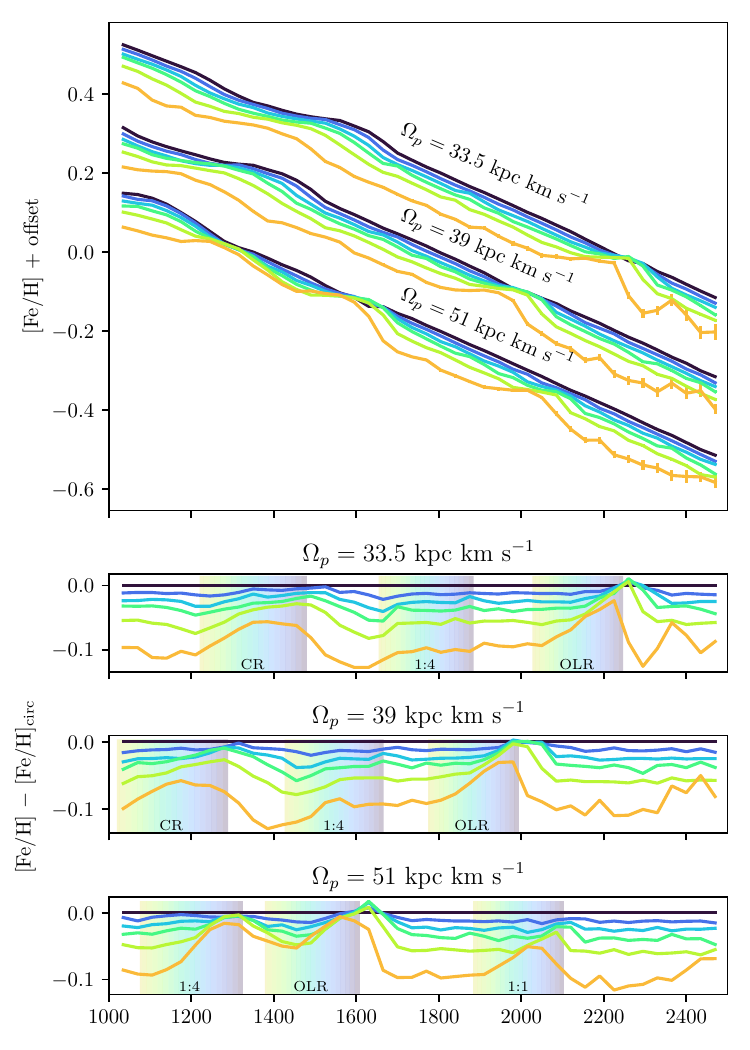}
    \caption{Mean [Fe/H] as a function of $J_\phi$ in bins (same as Figure \ref{fig:irene_explanation}) of $J_R$ for particles evolved in all three simulations.
    Error bars show the standard error.
    The qualitative flattening and wave-crest signatures are not sensitive to pattern speed.
    The vertical bands show the location of axisymmetric resonance lines, as in Figure \ref{fig:irene_explanation}.
    }
    \label{fig:irene_comparison}
\end{figure}

\begin{figure}
    \centering
    \includegraphics[width=0.45\textwidth]{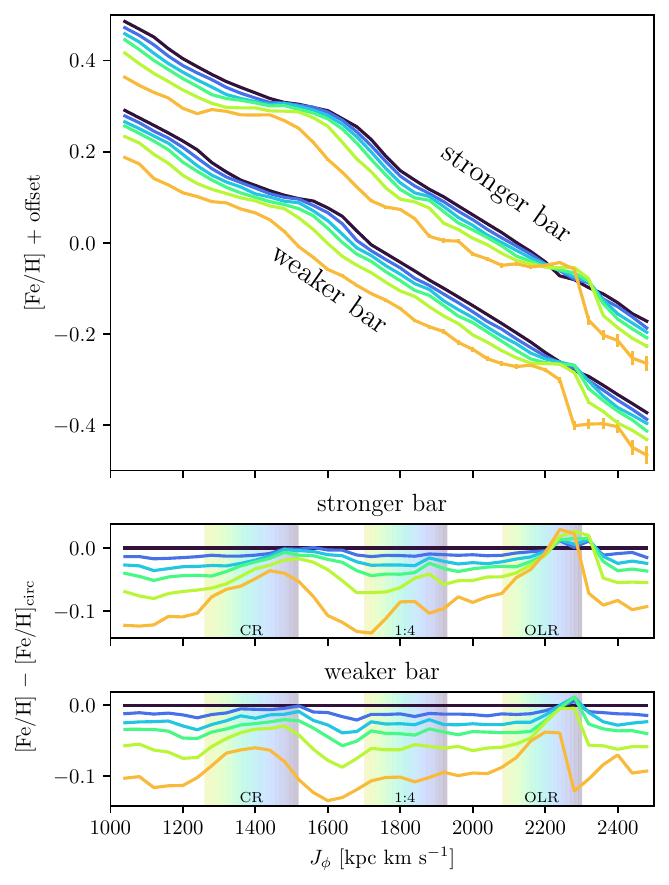}
    \caption{Mean [Fe/H] as a function of $J_\phi$ in bins (same as Figure \ref{fig:irene_explanation}) of $J_R$ for particles evolved with two different bar strengths, but the same pattern speed.
    Error bars show the standard error.
    \reviewer{Because the stronger bar traps more stars on resonant orbits, the metallicity signatures at CR and the OLR cover a larger range in $J_\phi$, and are therefore also stronger at larger $J_R$.}
    }
    \label{fig:barstrength}
\end{figure}

The qualitative wave-crest and flattening signatures at OLR and CR, respectively, in Figure \ref{fig:irene_explanation} do not change with pattern speed.
Figure \ref{fig:irene_comparison} shows the metallicity trends for particles evolved in all three simulations.
While the value of $J_\phi$ at which the resonances occur changes, the chemical signature of the resonances does not.
They do change with bar strength, however.
As the bar strength increases, the region of trapped orbits around the axisymmetric resonance line becomes larger (c.f. \citealt{binneyOrbitalToriNonaxisymmetric2018}).
Figure \ref{fig:barstrength} shows the flattening and wave-crest for the \slowbar{} simulation, and for stars evolved in the same potential but with a bar of twice the strength ($\alpha_{m=2} = 0.02$, $\alpha_{m=4} = -0.001$).
The flattening in [Fe/H] at CR as a function of $J_\phi$ is more prominent and extends across a larger range of $J_\phi$.
\reviewer{Likewise, the wave-crest at the OLR and 1:1 resonance becomes more extreme, with the mean metallicity for high-$J_R$ stars exceeding that of those on circular orbits in the resonance zone, and the effect extending across a wider range in $J_\phi$ for the hottest bins.}

\subsection{Action-metallicity signatures in the LAMOST data}

\begin{figure*}
    \centering
    \includegraphics[width=0.85\textwidth]{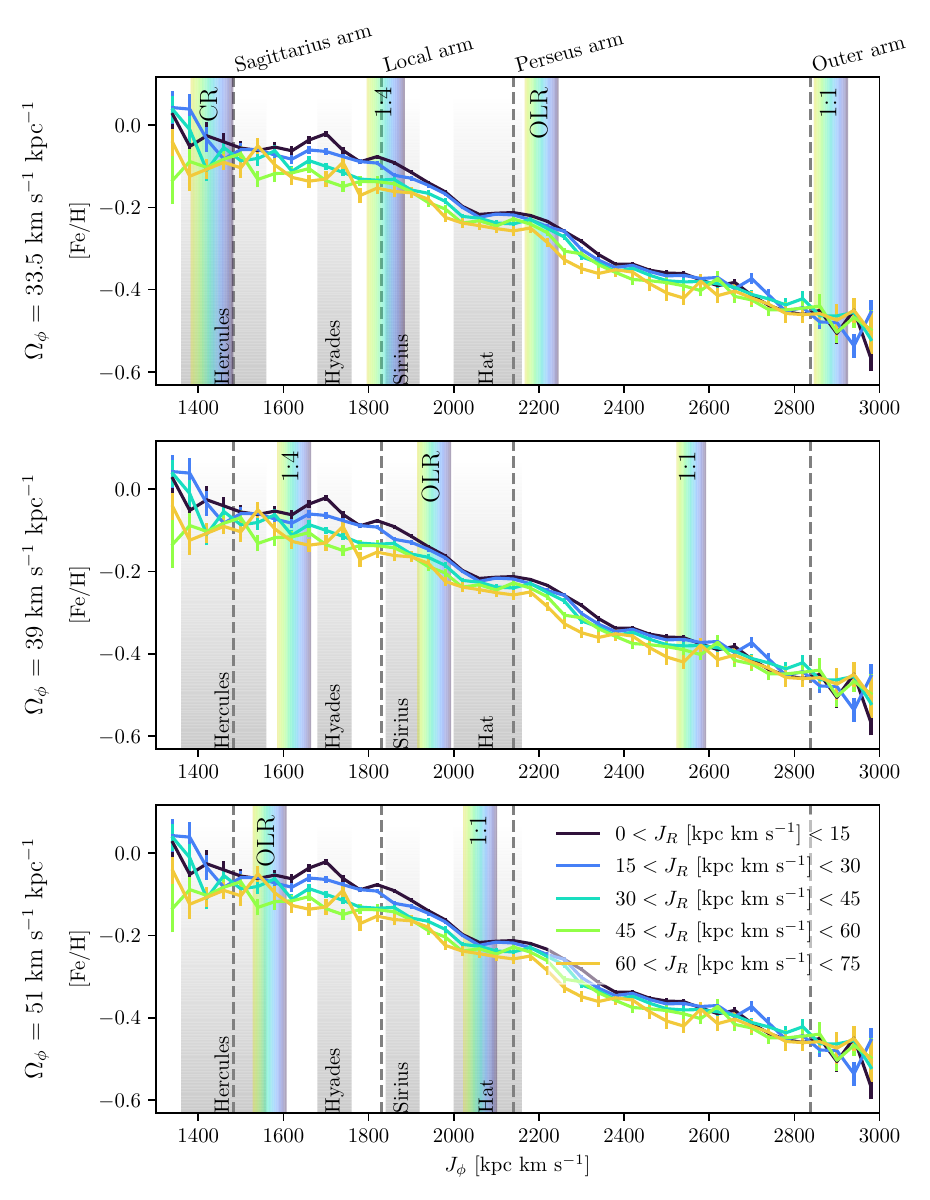}
    \caption{Mean [Fe/H] as a function of $J_\phi$ in bins of $J_R$ for \lamost{}.
    The errorbars show the standard error on the mean.
    There are not enough stars with $J_R > 60~\kpckms$ to plot trends.
    Grey vertical bands mark the approximate $J_\phi$ of known moving groups for orientation. 
    Gray dashed lines mark the $J_\phi$ values of circular orbits with $R$ at the radius where the solar azimuth intersects hypothesized spiral arms (taken from \citealp{reidTrigonometricParallaxesHigh2014}).
    Observational data is not yet plentiful enough to clearly identify or rule out the resonance signatures described in section \ref{sec:explanation}, but the flattening around $1400~\kpckms < J_\phi < 1700 ~\kpckms$ and the elbow at $J_\phi \approx 2200~\kpckms$ are suggestive of a slow bar.
}
    \label{fig:lamost_irene}
\end{figure*}

Figure \ref{fig:lamost_irene} shows mean [Fe/H] as a function of $J_\phi$ and $J_R$, the diagnostic introduced in section \ref{sec:explanation}, drawn with the observational data introduced in Section \ref{sec:data}.
The three panels show axisymmetric resonance line zones corresponding to the three pattern speeds used in the test particle simulations inspired by values in the literature. 
The limitations of available data (the number of stars observed, their footprint coverage, and systematics in their metallicities) make it difficult to unambiguously identify either the flattening or wave-crest signatures.
Only for $\Op = 33.5~\kmskpc$ are both the CR and the OLR region populated with data, although this is sensitive to the rotation curve of the approximate axisymmetric potential used to calculate actions.
Increasing incompleteness with distance from the Sun yields mean-[Fe/H] uncertainties large enough to obscure potential resonance signatures for $J_\phi \lesssim 1600~\kpckms$ and $J_\phi \gtrsim 2700~\kpckms$, even though some data is present.
As we will discuss in section \ref{sec:sellection}, the azimuthal coverage of the \lamost{} and \gaia{} data also obscure a potential signature of bar resonances.
\reviewer{Appendix \ref{appendix:apogee} briefly discusses the status of the wave-crest and flattening signatures in \apogee{}.}

In the context of the above caveats, we note that a possible wave-crest is present in Figure \ref{fig:lamost_irene} around $J_\phi \approx 2200~\kpckms$, i.e. close to the Hat moving group.
\reviewer{This corresponds to the OLR for $\Omega_p = 33.5~\kpckms$.
There is a flattening in the mean [Fe/H] trend at $1400~\kpckms \lesssim J_\phi \lesssim 1800~\kpckms$ ($6.5~\kpc \lesssim R \lesssim 8.55~\kpc$), which corresponds to CR for this pattern speed.}
Similar trends have been observed before in the Milky Way \citep{haydenCHEMICALCARTOGRAPHYAPOGEE2015, haywoodPhylogenyMilkyWay2018, haywoodRevisitingLongstandingPuzzles2019} and in other galaxies \citep{stottRelationshipSpecificStar2014, leethochawalitKeckSurveyGravitationallyLensed2016}, and have been largely attributed to well-mixed gas, although \citet{haydenCHEMICALCARTOGRAPHYAPOGEE2015} states that stellar migration due to the bar is the likely cause.
This flatting is mildly attenuated by our cuts in $\log g$ and $T_\mathrm{eff}$, suggesting that a more thorough understanding of metallicity systematics may shed light on it.
If the flattening is caused by stellar migration, rather than the birth properties of stars in the inner disk, it is consistent with the CR resonance of a slow (circa $33.5~\kpckms$) bar. 
\reviewer{At $J_\phi \approx 1850~\kpckms$, there is a small bump in the lowest $J_R$ bin which can be interpreted as corresponding to the 1:4 resonance for this pattern speed.}

\reviewer{
While the resonance locations for $\Omega_p = 33.5~\kmskpc$ are fairly consistent with the observational data, those for the other two pattern speeds are much less so.
For $\Omega_p = 39.5~\kmskpc$, neither the OLR no 1:4 resonance correspond to signature in the [Fe/H] trends, and the error bars at the 1:1 resonance are large enough to mask the weak signature expected there.
For $\Omega_p = 51~\kms$, the 1:1 resonance corresponds to a possible wave-crest, but the detailed trends at the OLR are masked by the error bars, and the flattening discussed above is not explained.
We conclude that a slow bar, ($\Omega_p = 33.5~\kmskpc$ or slightly faster) is most consistent with the observational data.
}

The dependence of mean [Fe/H] on $J_R$ differs strongly across $J_\phi$. 
For $J_\phi \lesssim 2000~\kpckms$ (at least until $J_\phi \approx 1600~\kpckms$, when the uncertainties become large), stars on higher-eccentricity orbits have lower [Fe/H].
For $J_\phi \gtrsim 2000~\kpckms$ the dependence is much weaker.
This effect is not a direct consequence of the dependence of [Fe/H] on height from the disk, $|z|$, surface gravity, $\log g$, or effective temperature, $\teff$.
It also does not appear to be caused by ``contamination'' from the high-$\alpha$ disk, since removing stars with large [Mg/Fe] (per \citealp{wheelerAbundancesMilkyWay2020}) does not have any effect.
Additionally, the dependence of mean [Fe/H] on $J_R$ exhibits the same behavior when Figure \ref{fig:lamost_irene} is made with \apogee{} \citep{majewskiApachePointObservatory2017} DR16 \citep{ahumada16thDataRelease2020}, rather than \lamost{} data (see Appendix \ref{appendix:apogee}).
We conclude that variation in correlation of mean [Fe/H] with $J_R$ is likely physical, rather than an artifact of the observational data or our analysis.


\section{Discussion} \label{sec:discussion}

\label{sec:sellection}
\begin{figure*}
    \centering
    \includegraphics[width=\textwidth]{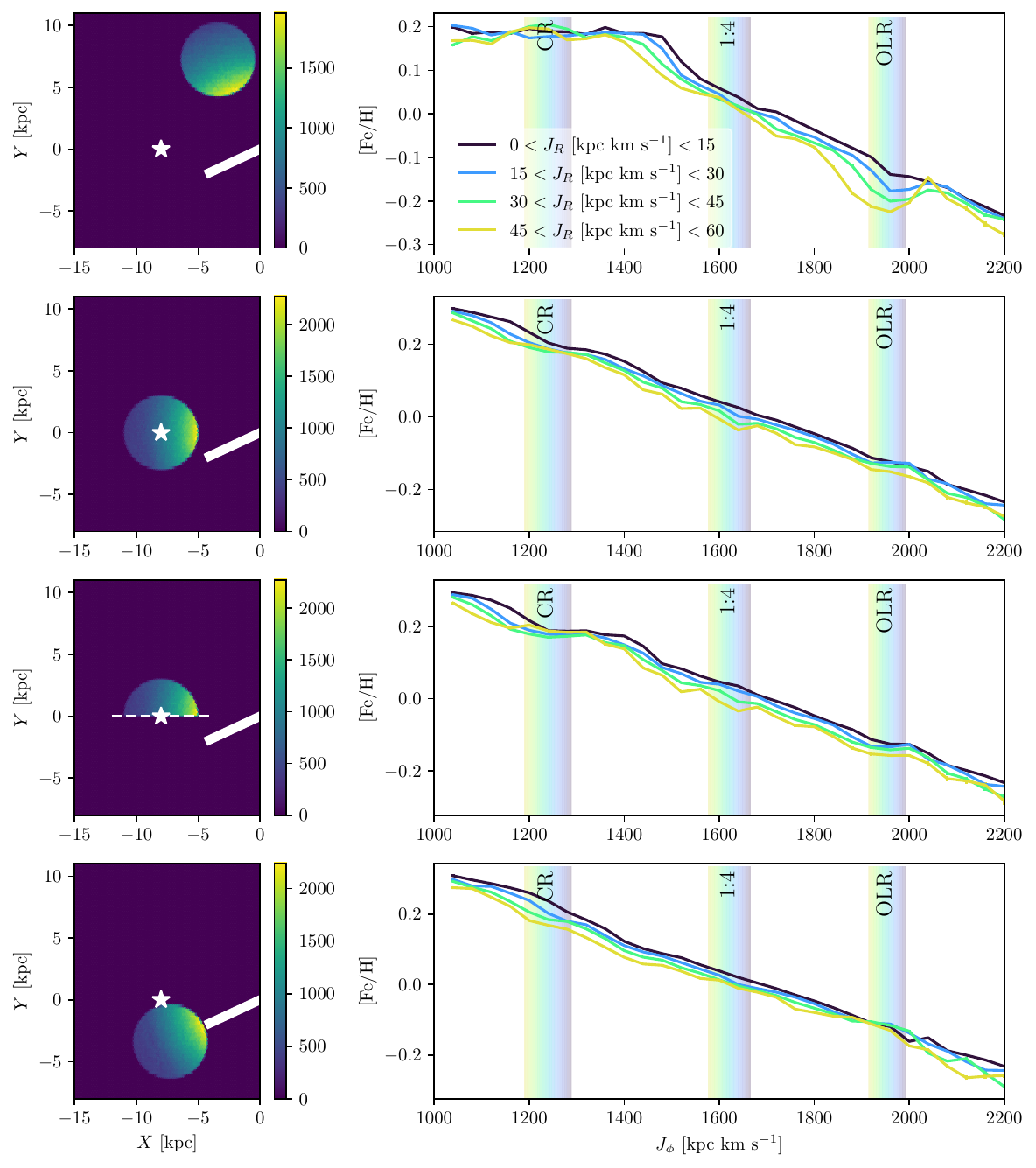}
    \caption{ The location of observed stars (i.e. the spatial selection function) has a strong impact on the flattening and wave-crest signatures.
    Here, resonance signatures as exhibited by stars in different azimuthal regions of the disk in the \intermediatebar{} simulation are plotted.
    $J_R$ bins above $60~\kpckms$ are not plotted because of the lack of sufficient data.
    Each row shows stars within a cylinder centered at the solar Galactocentric radius at different azimuths: exactly along the minor axis of the bar (top), in the position of the Sun (middle rows), and exactly along the bar's major axis (bottom).
    The left column shows number density in the $X$--$Y$ plane, i.e. the footprint of each mock data set in the disk.
    The white line marks the orientation of the bar, and the white star marks the location of the Sun.
    The right column shows metallicity trends in $J_\phi$ and $J_R$ for each region.
    In row 3, only stars with azimuth farther from the bar azimuth than the Sun are used.
    The CR flattening and the OLR wave-crest are much more prominent in stars anti-aligned with the bar, but could be seen in a large volume of local data by plotting only stars farther from the bar's azimuth than the Sun (row 3). 
    }
    \label{fig:simple_sellections}
\end{figure*}

The bar and spiral arms have been shown to lead to a nonaxisymmetric metallicity distribution (e.g. \citealp{dimatteoSignaturesRadialMigration2013, grandSpiralinducedVelocityMetallicity2016, khoperskovStellarMetallicityVariations2018, fragkoudiDiscOriginMilky2018}), meaning that the spatial location of the observational data with respect to the bar has a significant impact on both the wave-crest and flattening signatures.
In contrast to previous Figures, which use stars at all azimuths, Figure \ref{fig:simple_sellections} shows mean [Fe/H] trends in $J_\phi$ and $J_R$ using only stars within cylinders centered at different points in the disk for the \intermediatebar{} test particle simulation, demonstrating that stars with azimuth nearer to the bar's minor axis show the resonance signature much more prominently.\footnote{Particles with positive $X$ are rotated $180^\circ$ to the Sun's side of the Galaxy to increase the number available to plot.}
All cylinders are centered on points 8 kpc from the center of the disk, the approximate Galactocentric radius of the Sun.
The first row, which plots stars centered on a point on the bar's minor axis (i.e. with $\phi = \phi_\odot - 75^\circ$, taking the bar's pitch angle to be $25^\circ$), exhibits resonance signatures much stronger than the others.
For these stars, the flattening at CR turns sharply on and off for each $J_R$ bin, and the OLR wave-crest is very strong, with mean [Fe/H] dependence on $J_R$ fully inverting at $J_\phi \approx 2250~\kpckms$.
For survey volumes centered on the Sun or aligned with the bar's major axis (rows 2 and 4 in Figure \ref{fig:simple_sellections}), the CR flattening and OLR wave-crest are attenuated and nearly absent, respectively.
But if only the stars within 3 kpc of the Sun which are farther from the bar azimuth than the Sun are plotted, both signatures become much more prominent (Figure \ref{fig:simple_sellections} row 3), demonstrating that a sample in the ``extended solar neighborhood'' can be used to uncover these features.

\begin{figure}
    \centering
    \includegraphics[width=0.45\textwidth]{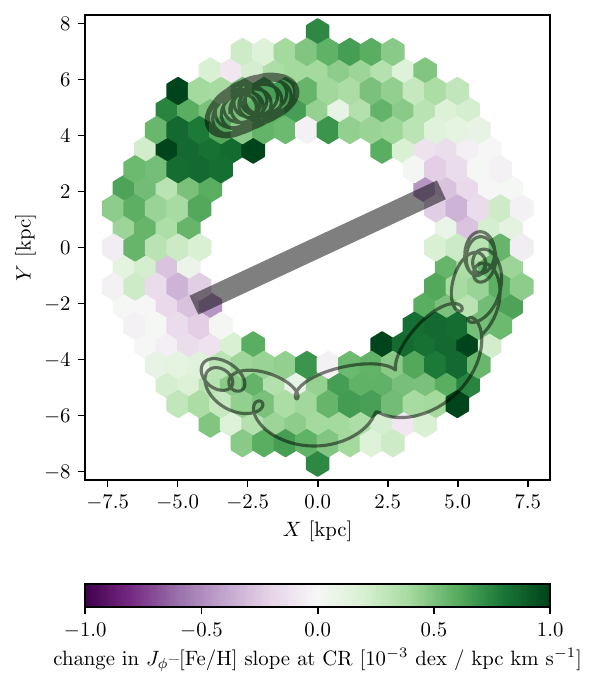}
    \caption{
    The ``banana-shaped'' orbits of stars trapped at CR spend little time near the bar's major axis, and many dwell exclusively near its minor axis, which explains the azimuthal sensitivity demonstrated in Figure \ref{fig:simple_sellections}.
    The relative slope of the $J_\phi$--[Fe/H] relation around CR with respect to the overall slope for stars with $J_R < 60~\kpckms$ as a function of position in the disk for the \intermediatebar{} simulation.
    The black rectangle indicates the orientation of the bar.
    The orbits of two stars trapped by the CR resonance are overplotted (in the rotating frame).  
}
    \label{fig:crmap}
\end{figure}
Figure \ref{fig:crmap} shows the degree of CR flattening as a function of position in the disk with two CR orbits plotted, demonstrating again the importance of the azimuth of observed stars and helping to reveal its cause.
It is colored by the slope of the $J_\phi$--[Fe/H] relation for stars roughly in the \intermediatebar{} CR region, with $1200 ~\kpckms < J_\phi < 1300~\kpckms$ and $J_R < 60~\kpckms$, minus the slope outside the CR zone (calculated with the stars in $1600~\kpckms < J_\phi < 1800~\kpckms$). 
This figure shows, similarly to Figure \ref{fig:simple_sellections}, that the CR flattening is much stronger in stars around the bar's minor axis.\footnote{There is a weak asymmetry in the metallicity slope at CR around the bar's minor axis, at about $Y = 4~\kpc$, slightly ahead of the minor axis in Figure \ref{fig:crmap}. This is a phase-mixing relic in the simulation which depends on how stars with different initial phases and metallicities get scattered at CR.}
The two grey lines show the paths of two stars trapped by the CR resonance as they orbit over 3000 time steps.
The azimuthal dependence is a consequence of the ``banana'' shape of these orbits, which oscillate around Lagrange points 4 and 5 on the bar's minor axis. 

\begin{figure*}
    \centering
    \includegraphics[width=\textwidth]{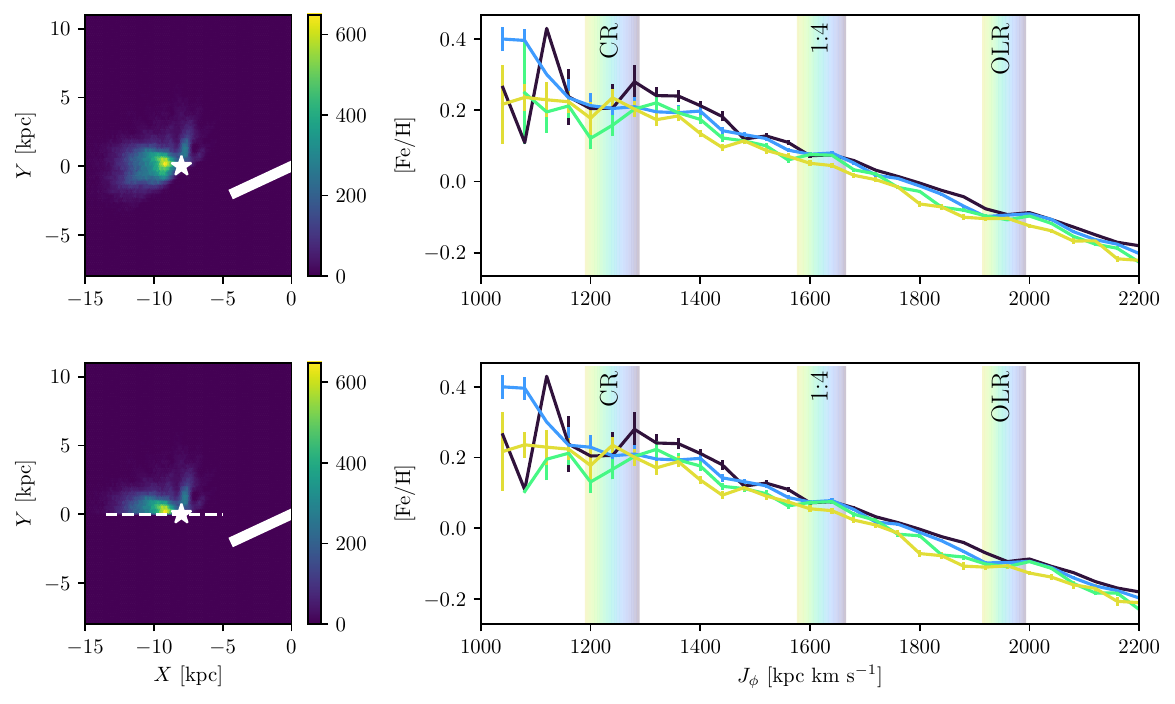}
    \caption{Mean [Fe/H] as a function of $J_\phi$ and $J_R$ for the \intermediatebar{} simulation with the \lamost{}'s spatial selection function imposed. 
    \textbf{top: }full \lamost{} footprint.
    \textbf{bottom: }partial \lamost{} footprint, with only stars with azimuth less than the Sun's.
    The flattening and wave-crest signatures at CR and the OLR, respectively, are extremely attenuated because stars are concentrated near the Sun. 
    }
    \label{fig:lamost_selections}
\end{figure*}

We use a simple rejection sampling scheme to simulate the \lamost{} spatial selection function in detail and understand why current data is inadequate to reveal the signatures explored in this work.
We approximate the spatial distribution of the test particles and \lamost{} stars, $p_\mathrm{sim}(\mathbf{r})$ and $p_\mathrm{LAMOST}(\mathbf{r})$, where $\mathbf{r}$ is 3D position, by counting the fraction of stars in each bin in a fine grid\footnote{Bin sizes are 0.5 kpc in the $X$ and $Y$ dimensions, and 0.1 kpc in the $z$ dimension}, then interpolating with a cubic spline.
We then include each test particle in the subset for comparison with probability proportional to the ratio of $p_\mathrm{sim}(\mathbf{r})/p_\mathrm{LAMOST}(\mathbf{r})$.
This reduces the number of particles from $4\times 10^7$ to about $1.6 \times 10^5$, but allows a more ``apples to apples'' comparison between simulation and data.
Figure \ref{fig:lamost_selections} shows the mean [Fe/H] trends in action space for the \intermediatebar{} simulation with the observational spatial footprint imposed. 
(The results are similar for the \slowbar{} and \fastbar{} simulations.)
Both the wave-crest and flattening are attenuated and distorted by the limited azimuthal coverage.
The first row uses the full observational footprint, and the second row uses the same cut as Figure \ref{fig:simple_sellections} row 3, keeping only stars with azimuth closer to the bar's minor axis than the Sun.
Both selections are too concentrated near the Sun to show the signatures clearly.
While \lamost{} and other current datasets don't have the spatial coverage to look for resonance signatures in stars further away from the bar's major axis, in the near-future metallicities from the \gaia{} radial velocity spectrograph or Sloan-V \citep{kollmeierSDSSVPioneeringPanoptic2017} will likely make such an analysis possible.

\section{Conclusions} \label{sec:conclusions}
We investigate the effect of bar resonances on the metallicity distribution in the Milky Way's thin disk as a function of the stars' orbital actions, $J_\phi$ and $J_R$.
We use test-particle simulations of a stellar disk in a Milky-Way-like potential with a simple rotating bar to identify distinct signatures: the corotation resonance leads to a flattening of the metallicity gradient, and higher-order resonances, especially the Outer Lindblad resonance; this creates a ``wave-crest'' in the mean [Fe/H] as a function of $J_\phi$ binned by $J_R$. 
We confirm that these signatures do not change qualitatively with pattern speed or strength, although the strength of the signal varies with bar strength.
We demonstrate that a survey's selection function can strongly affect the appearance of these signatures.

We search for these resonant metallicity-vs.-action features in stars using metallicities and radial velocities from \lamost{} DR 5 and parallaxes, proper motions, and supplementary radial velocities from \gaia{} EDR3.
While distinguishing signatures of bar and spiral resonances remains difficult, the diagnostics discussed here can identify both the OLR and CR resonance at once.
We find weak evidence for a slow bar with an OLR associated with the ``hat'' moving group, but conclude that present data does not allow us yet to unambiguously identify these signatures.
\secondrevision{We do not know how strong the real bar in the MW is, and also do not know if and by how much the quoted measurement uncertainties are underestimated.
However, even by exploring different combinations, we did not manage to produce an exact match of the model to the data.
This could indicate that additional mechanisms might be at play in the Galactic disk (pattern speed variations, spiral arms, other mixing mechanisms. etc.) or that the relationship between metallicity and birth radius is not captured will by either of the models we used.
But the investigation of their exact influence on the bar’s action-metallicity pattern is beyond the scope of this paper.}

Interpreting observational data is made more difficult, not only by the complexities of the Galaxy not present in our simplified test particle model, but also because current surveys provide 6-D kinematics and metallicities only for stars in a limited region of the Galactic disk.
In particular, we show that the strength of the flattening and wave-crest signatures depends on the Galactocentric azimuth.
The strongest signature is found at an azimuth along the bar's minor axis, while along the bar's major axis the resonant signature in the metallicity almost vanishes.
Consequently, our observing location within the Galactic disk, $\sim 25^\circ$ behind the bar, is not ideal, but observations of more stars with Galactocentric azimuth far from the major-axis of the bar, which near-future data will deliver, can remedy this.

There are several promising directions for future progress on chemo-dynamically identifying bar resonances.
In this paper, we have used a fairly simple galaxy model as basis for the comparison to the data. More complex models may yield modified predictions. 
For example, a bar with significant deceleration will trap stars with a different distribution of orbits \citep{chibaResonanceSweepingDecelerating2019}, and \citet{chibaTreeringStructureGalactic2021} show that metallicity should decrease from the center of the corotation resonance to its boundary.
$N$-body models with self-consistent bars could also provide a richer picture, including the possibly confounding effects spiral arms, which impact metallicity distributions and can overlap with bar resonances \secondrevision{(e.g. \citealp{khoperskovChemokinematicsMilkyWay2021, asanoImpactBarResonances2022})}. 
Stellar ages and chemical abundances beyond [Fe/H] provide additional clues to a star's birth location and, given a better understanding of the Galaxy's chemical evolution, can doubtlessly further constrain birth radius.
On the other hand, if the bar resonances can be conclusively identified by other means, their chemical abundances inform us on the star formation history of the disk.
Sloan-V and \gaia{}'s future data releases, which will increase the amount of suitable data and its azimuthal coverage, will likely allow us to pinpoint the pattern speed of the bar and shed light on the Galaxy's dynamical structure. 

\software{galpy \citep{bovyGalpyPythonLibrary2015}, Matplotlib \citep{hunterMatplotlib2DGraphics2007}}

\section*{Acknowledgments}
The authors would like to thank Jason Hunt, Tomer Yavetz, Kathryn Johnston, and the Milky Way stars group at Columbia for useful discussion and suggestions.

AJW is supported by the National Science Foundation Graduate Research Fellowship under Grant No. 1644869. MKN is in part supported by a Sloan Research Fellowship.

Guoshoujing Telescope (the Large Sky Area Multi-Object Fiber Spectroscopic Telescope LAMOST) is a National Major Scientific Project built by the Chinese Academy of Sciences. Funding for the project has been provided by the National Development and Reform Commission. LAMOST is operated and managed by the National Astronomical Observatories, Chinese Academy of Sciences. 

This work has made use of data from the European Space Agency (ESA) mission {\it Gaia} (\url{https://www.cosmos.esa.int/gaia}), processed by the {\it Gaia}
Data Processing and Analysis Consortium (DPAC, \url{https://www.cosmos.esa.int/web/gaia/dpac/consortium}). Funding for the DPAC has been provided by national institutions, in particular the institutions participating in the {\it Gaia} Multilateral Agreement.

\bibliography{bdc}

\appendix

\section{Simulated observational error} \label{appendix:fuzz}
\begin{figure}
    \centering
    \includegraphics[width=0.95\textwidth]{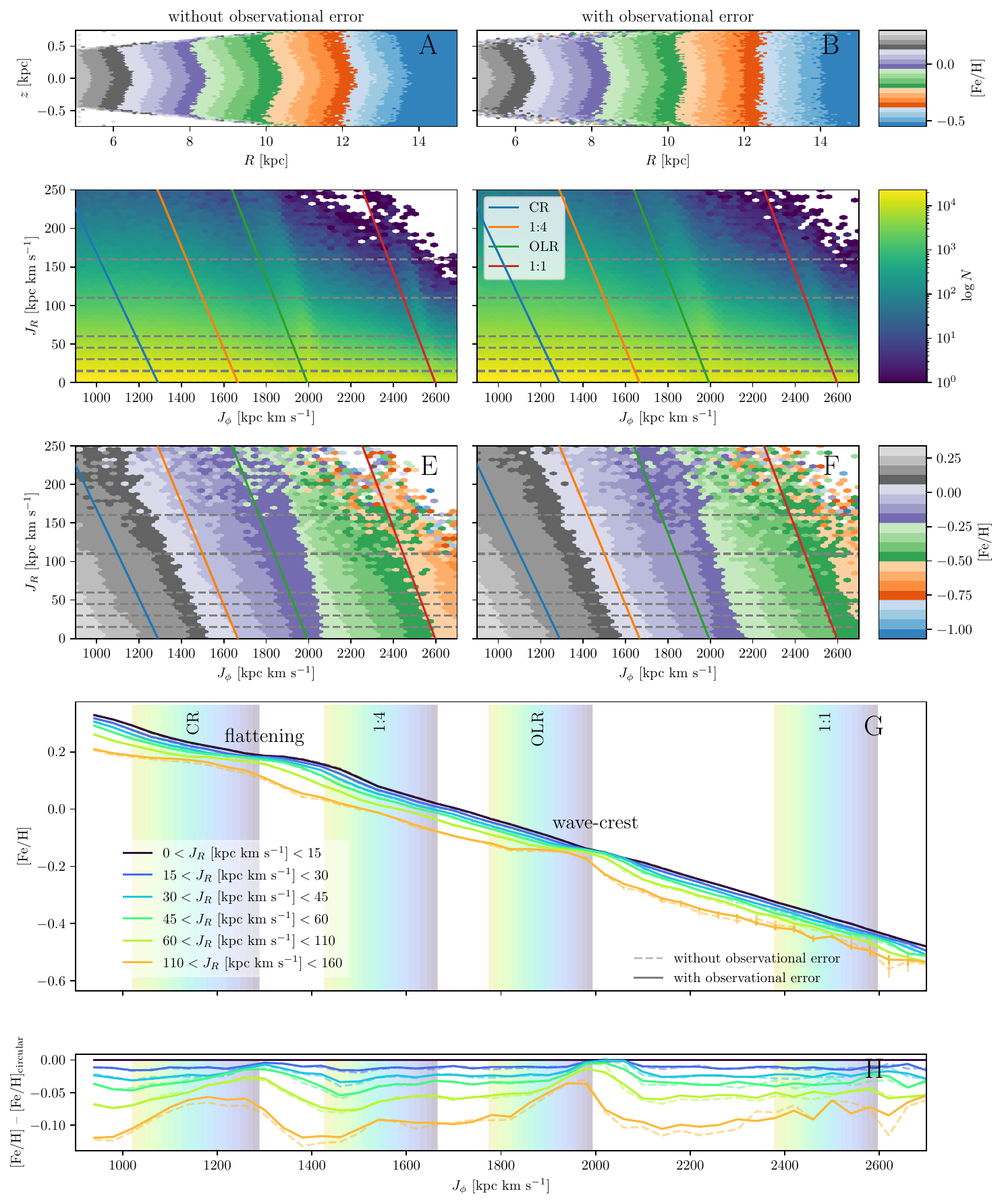}
    \caption{Same as Figure \ref{fig:irene_explanation}, but with simulated observational radial velocity and distance error applied. The left column shows the unperturbed simulation and the right column shows the perturbed one.  Both are barred. Measurement error on radial velocity and distance has only has very small effect on the qualitative signatures discussed in this paper.}
    \label{fig:fuzzed_explanation}
\end{figure}

While we do not fuzz our model with simulated observational error, doing would not effect our results.
Figure \ref{fig:fuzzed_explanation} compares the resonance signatures computed for a version of the \intermediatebar{} test-particle simulation with and without an radial velocity error of $3~\kms$ and distance error of 10\% (roughly the mean observational values).
The difference is minimal, with the largest changes being in the meridional plane, and thus having no impact on our proposed signatures.
Because kinematic errors perturb calculated $J_R$, the metallicity contours in the $J_\phi$--$J_R$ plane are slightly steeper when observational error is applied.
\secondrevision{Though the expected measurement errors have little impact, they may wash out the resonance signatures if significantly underestimated.}

\section{Axisymmetric approximate actions calculated with the Eilers 2019 potential}\label{appendix:eilers}
\begin{figure}
    \centering
    \includegraphics[width=0.8\textwidth]{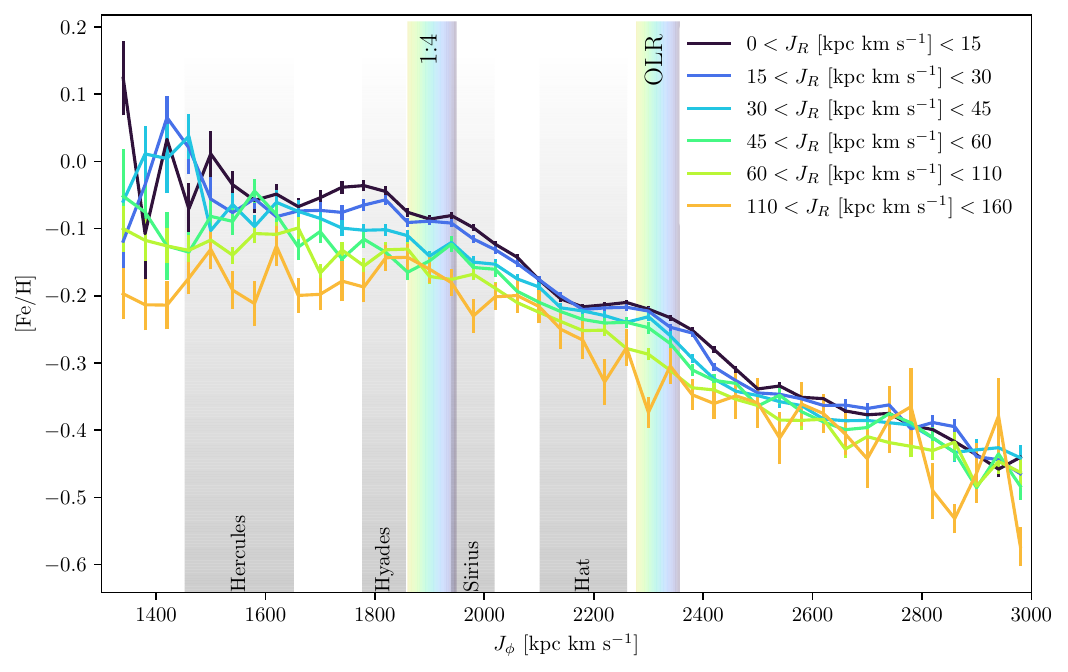}
    \caption{LAMOST data just as in Figure \ref{fig:lamost_irene}, but with actions calculated using the \citet{eilersCircularVelocityCurve2019} potential. The overplotted resonances are for $\Omega_\mathrm{p} = 36~\kpckms$.  The CR and 1:1 resonances are not within the plotted $J_\phi$ range. 
    The use of this potential has no appreciable impact on the interpretation of the observational data.
    }
    \label{fig:obs_eilers}
\end{figure}

To better understand the dependence of our results on the potential used to calculated axisymmetric actions, in Figure \ref{fig:obs_eilers} we show our main observational plot with actions calculated using the \citet{eilersCircularVelocityCurve2019} potential, which has $R_0 = 8.112~\kpc$ \citep{gravitycollaborationDetectionGravitationalRedshift2018} and $v_\mathrm{circ}(R_0) = 229~\kms$.
While overall action ``scale'' of the resonances locations and metallicity trends are sensitive to this choice, their relative positions and detailed shapes are much less so.
Axisymmetric resonance lines are plotted for $\Omega_\mathrm{p} = 36~\kpckms$, which corresponds to the Hat moving group, as the \slowbar{} pattern speed does when using \texttt{MWPotential2014}.

\section{Alternate metallicity assignment} \label{appendix:alternate_metallicity}
\begin{figure}
    \centering
    \includegraphics[width=\textwidth]{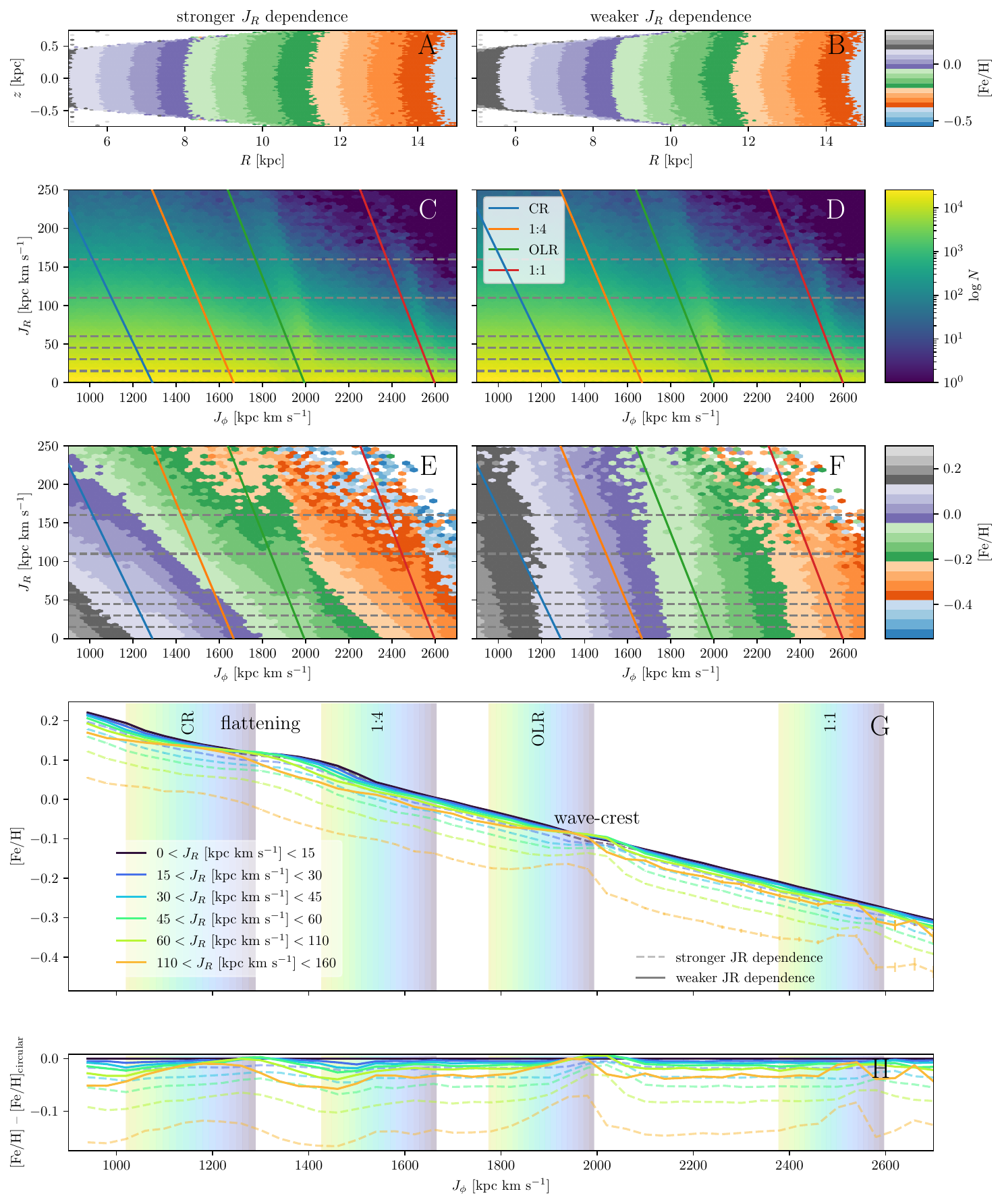}
    \caption{Similar to Figure \ref{fig:irene_explanation}, but comparing the \intermediatebar{} simulation with metallicities painted on according to initial action with two different dependencies on $J_R$ (left and right column, dashed and solid lines, respectively). See text for details. Our proposed signatures are largely the same when metallicities are assigned this way.}
    \label{fig:altz}
\end{figure}

\citet{debattistaBoxPeanutshapedBulges2020} found, in the context of the Galactic bulge, that metallicities assigned to star particles in simulations are more accurate when painted on according to initial axisymmetric action.
Figure \ref{fig:altz} shows our proposed signatures when metallicities are assigned according to 
\begin{equation}
    \mathrm{[Fe/H}] = \frac{0.003}{\kpckms} J_{\phi,0} - \frac{\alpha}{\kpckms} J_{R,0} + 0.5 + \epsilon \quad,
\end{equation}
with $\alpha = 0.003$ (right column), which approximately matches the radial metallicity trends in Figure \ref{fig:lamost_kinematics}, and with $\alpha = 0.012$, which more closely matches the procedure in \citet{debattistaBoxPeanutshapedBulges2020} (left column). 
We have not included a $J_z$ dependence, since we restrict the data to a small range in $J_z$.
In both cases, the signatures of resonances are very similar to what we obtain when assigning metallicity based on initial position.
In the $\alpha = 0.12$ case, the resonance signature are ``softened'', since the the mean [Fe/H] trends for each $J_R$ bin are further apart, but at the same time the amplitude of the signature in each [Fe/H] trend is slightly amplified.

\section{APOGEE data} \label{appendix:apogee}

\begin{figure}
    \centering
    \includegraphics[width=\textwidth]{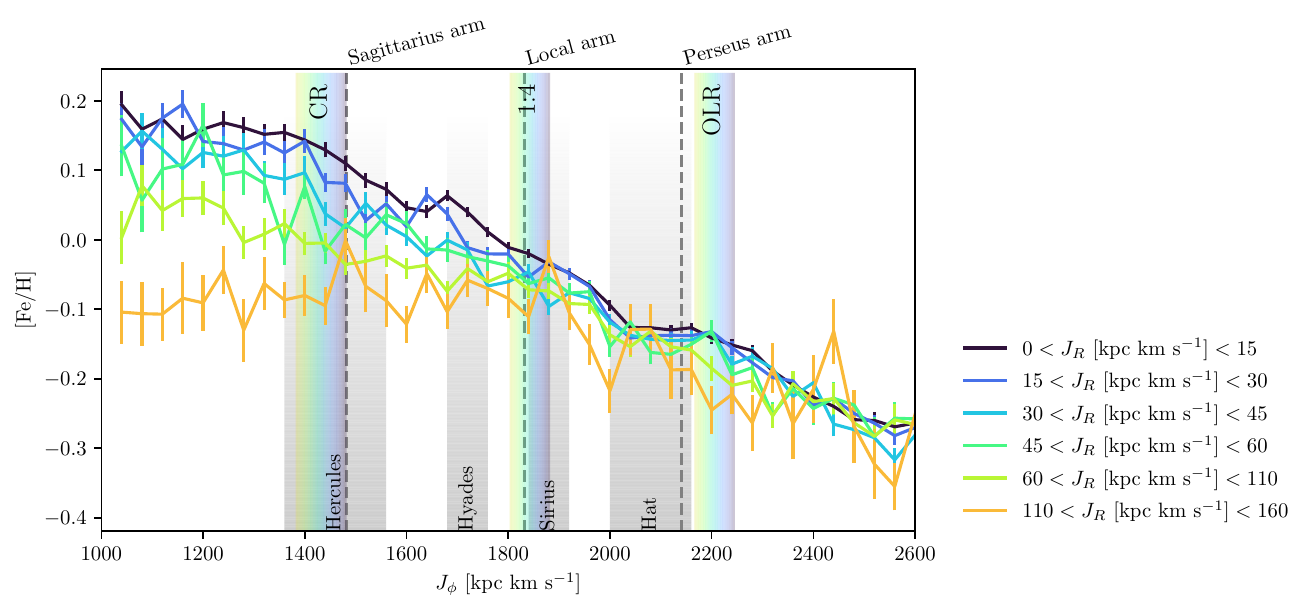}
    \caption{Same as the top panel ($\Omega_p = 33.5$ \kpckms) in Figure \ref{fig:lamost_irene}, but using \apogee{} DR16 instead of LAMOST. Note that \apogee{} observes more stars towards the Galactic center than LAMOST and populates also $J_\phi \lesssim 1300\kmskpc$. 
    The smaller number of stars (compared to \lamost{}) means that it's not possible to draw strong conclusions from the observational data.}
    \label{fig:apogee}
\end{figure}

Figure \ref{fig:apogee} is a version of our main observational plot using \apogee{} DR16 \citep{gunnTelescopeSloanDigital2006, blantonSloanDigitalSky2017, majewskiApachePointObservatory2017, wilsonApachePointObservatory2019, ahumada16thDataRelease2020} (with the cuts described in Section \ref{sec:data} applied).
The error bars are too large to support the identification or ruling-out of the wave-crest and flattening signatures, but hints thereof can be seen in light of the \lamost{} data.
\reviewer{Interestingly, the flatting in the inner Galaxy noted in \citet{haydenChemicalCartographyAPOGEE2014} and others is mildly attenuated by our cuts, and is more present in high-$J_R$ bins.}

\end{document}